# Empirical performance evaluation of Enhanced throughput schemes of IEEE802.11 technology in Wireless area Networks

Femi-Jemilohun Oladunni .Juliet, Walker Stuart

School Of Computer Science and Electronic Engineering University of Essex
Colchester, Essex, United Kingdom
`{ojfemi and stuwal}@essex.ac.uk`

## ABSTRACT

*The success in the growing wireless standards can be measured by the achievement of quality of service (QoS) specifications by the designers. The IEEE802.11 wireless standards are widely accepted as wireless technology for wireless LAN. Efforts have been made over the years by the task group to provide adequate number of QoS enhancement schemes for the increasing numbers of multimedia applications. This paper examines the empirical performances of ad hoc wireless networks deployed on IEEE802.11 standard variants. A survey to some of the QoS schemes incorporated in IEEE802.11 wireless PHY layers were carried out. Then the effects of this enhancement schemes in relation to data throughput and system capacity and reliability in the newest technology deployed on IEEE802.11ac standards was investigated using real time applications and simulation based approaches.*

## 1. INTRODUCTION

The accessibility to simple, flexible, scalable, cheap and ubiquitous communication provided by IEE802.11 standards had qualified it to be the major wireless technology all over the world. The wave of spectrum licence in the microwaves bands has led to growth explosion in the multimedia applications in the recent times hence; the demand for high throughput, streaming capability in video audio, and web services on the go. The high quality of services required by these applications, such as secured bandwidth, minimal data transfer delay spread, reduced jitter and error rate, have been a huge challenge for IEEE802.11 standards group. The experience in the wireless networks has been that of overcrowding poor quality of service, system degradation, as well as co-channel interference among the co-existing access networks. [1].

In this work, the experiences of the networks deployed on the standards in the real world scenario are being looked into. These standards are operated on the license free microwaves bands of the frequency spectrum. Some works have been carried out to investigate the system performance in IEEE802.11 wireless network systems, but the latest research works on the shielding enhancement techniques for high throughput delivery was not included. Therefore, this work intends to give a broad overview study of the enhancements schemes and their effects as determined by the networks delivery performances based on the coverage areas, received signals, interference between channels, and throughput capability in the IEEE802.11standards variants.

The rest of this work is organised as follows: Section two gives the overview of the related published work on IEEE802.11 WLAN performances. The background study of IEEE802.11 and the various shielding techniques for quality of service enhancement are addressed in section three. The empirical investigations of co-located networks performances of IEEE802.11 standards deployed on microwave bands and the newest technology: IEEE802.11ac, and development and implementation of beamforming algorithm were reported in section four. The empirical models of the measurement results of the signal strength distribution of the networks on microwave bands, the upcoming technology: IEEE802.11ac and discussions were presented in section five. The conclusion was given in section six.
.

## 2. RELATED WORKS.

In literature, there are some works on System performance evaluation and interference mitigation techniques in WLAN deployed on IEEE802.11 standards. However, very few of these authors have engaged physical devices in their works for real time applications assessment. Some of these are highlighted below.

### 2.1. Related works of WLAN indoor coverage.

Different techniques have been adopted in the previous works to achieve the target. Some of these techniques are: continuous wave transmitter with power meter receivers. [2] used a theoretical hybrid model that combined a two-dimensional ray tracing model with a direct transmitted ray (DTR) model to predict the radio coverage on single floors in multifloored buildings. Another technique is Broadband Pulse channel sounding. [3], measured the pulse response of network at 2.4GHz inside a building, he modulated the carrier with a repetitive 5ns pulses and transmitted it through a wideband biconical antenna. Also signal strength percentage reported by wireless LAN card, [4] used measured data and empirical models to predict the propagation path loss in and around residential area for the UNII band at 5.85GHz. While [5] developed throughput prediction model using the available measurement software products called SiteSpy and LanFielder to measure wireless LAN throughput and other network performances criteria and recorded the results in a precise site-specific manner. His measurement was limited to IEEE802.11b/n wireless system. [6] examined the impact of structural wall shielding on the system performance of an interference limited CDMA system using ray-tracing propagation models based on Geometric Optics and Uniform Theory of Diffraction to estimate the received signal. They observed a logarithmic relationship between the average output probability and the extent of a shielding in a typical single-floor office environment. They concluded that efficient deployment of shielding will enhance both system capacity and signal strength of wireless system. [7], in their work employed analytical approach using the Okumura's model to determine the propagation range of IEEE802.11 Radio LAN's for outdoor applications. This work used a lot of assumptions and does not reveal the performance range in the real application scenario. To evaluate the performance of 802.11b, [8] used two laptops equipped with Orinoco WLAN cards, and two PCs with D-Link WLAN cards. The combination of equipment from different chipset manufacturer may not favour adequate reading required. In [9] Discrete-event simulation of MAC portion of the IEEE802.11 protocol was used to evaluate the performance of the standard. [10-12] also used empirical methods to show the coverage area and the signal strength inside the coverage area to reveal the technology that gives better WLAN. [13] using computer simulation of a typical 802.11b/e access to an IP core network through an access point in an infrastructure WLAN.in the same manner, [14] conducted field performance comparison between IEEE802.11b and power line based communication network. They concluded that power line network outperforms IEEE802 .11b. [15-18] is similar in some aspects to what is done in this work. They used the experiment to validate the analytical performance models for IEEE802.11b .It was noted in their results that analytical result estimate experimental data with a mean error of 3-5%. In [19] Primary-Secondary transmission scheme was used to mitigate interference by beam forming and increase system capacity in WLAN systems where matlab simulation was engaged. [20] Investigated the efficiency of multi-cast beamforming optimization over IEEE802.11n WLAN. [21] Proposed the combination of advanced distributed beamforming techniques at physical layer for the improvement of the overall network capacity. Analytical modelling to evaluate the performance of ad hoc networks with M-element array was conducted by [22]. The upper and lower bounds on the transmission capacity in the multiple antenna ad hoc networks with multi-stream transmission and interference cancellation at the receiver was derived by [23] to increase system capacity through beamforming technique. Similar to our work, [24] gave the overview of the newest technologies promised to deliver multi-gigabits throughput: Ieee802.11ac/ ad, they described the channelization PHY design, and MAC modifications and beamforming in the standards.[25] proposed a joint adaptive beamforming algorithm for interference suppression in GNSS( Global Navigation Satellite System) receivers: first the orthogonal projection is used to suppress strong interference by producing nulling in the array pattern, while the maximum post-correlated C/No constrained is used to process the interference-free signal to further enhance the signal quality. [26] designed an advance interference resilient scheme for asynchronous slow frequency hopping wireless personal area networking and time division multiple access cellular system for interference reduction in wireless system. Different approaches have been engaged by different authors in their works to analyse the performances of wireless networks deployed on the IEEE802.11 technology but none has employed real time application approach, especially in the newest technology deployed in IEEE802.11ac for the empirical performances assessment. Hence this work differs from all in this aspect.

## 3. IEEE802.11

This is a set of standards for implementing WLAN (wireless local area networking) communications in the 2.4, 3.6, and 5 GHz frequency bands. The first in this group is IEEE80.11-2007 which has been subsequently amended. Other members are 802.11b and 802.11g protocols. These standards provide the basis for wireless network products using the Wi-Fi brand. IEEE 802.11b and g use the 2.4 GHz ISM band and are operated on WiFi channels. WiFi channel are grouped into 14 overlapping channels. Each channel has a spectral bandwidth of 22MHz, (though the nominal figures of the bandwidth of 20MHz are often given). This channel bandwidth exists for all standards even though different speed is available for each standard: 802.11g standard has 1, 2, 5, or 11 Mbps while 802.11g standard has 54Mbps. The difference in speed depends on the RF modulation scheme used. The adjacent channels are separated by 5MHz with the exception of 14 with the centre frequency separated from channel 13 by 12MHZ. From figure 1, it is obvious that a transmitting station can only use the fourth or the fifth channel to avoid overlapping. Most often, Wifi routers are set to channel 6 as the default, hence channels 1, 6, and11 have been adopted generally and particularly in Europe being non-overlapping channels for wireless transmission in the ISM band. This band can provide up to 11Mbps [8-10]

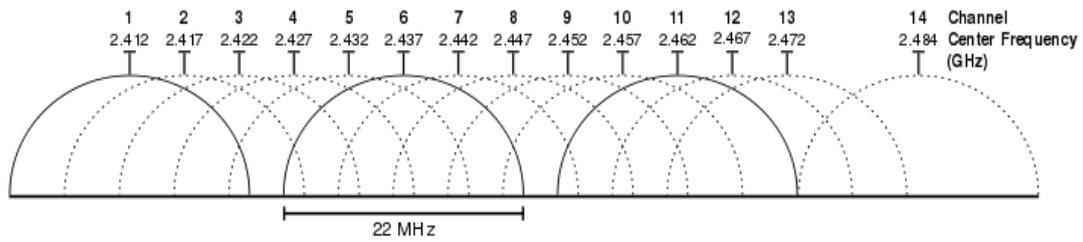

Figure 1. Structure of Wi-Fi Channels in the 2.4 GHz band [11, 12]

The UNII (unlicensed national information infrastructure) band is allocated for two radio devices: the unlicensed part is classified into A (5150-5350) MHz and B (5470-5725) MHz; the higher band which is licensed is C (5725-5859) MHz used in the installation of Fixed wireless Access (FWA) services, while A is for indoor mobile and B for indoor and outdoor WLANs. 5GHz U-NII bands which offers 23 non-overlapping channels, can provide up to 54Mbps in the WLAN and it is deployed in the IEEE802.11a/n standards.[11, 12]

Table 1. Wi-Fi channels in the 5GHz[27]

| Channel | Frequency /GHz | Band(A,B) | Maximum EIRP |
|---|---|---|---|
| 36 | 5.18 | A | 200mW |
| 40 | 5.20 | A | 200mW |
| 44 | 5.22 | A | 200mW |
| 48 | 5.24 | A | 200mW |
| 52 | 5.26 | A | 200mW |
| 56 | 5.28 | A | 200mW |
| 60 | 5.30 | A | 200mW |
| 64 | 5.32 | A | 200mW |
| 100 | 5.50 | B | 1W |
| 104 | 5.52 | B | 1W |
| 108 | 5.54 | B | 1W |
| 112 | 5.56 | B | 1W |
| 116 | 5.58 | B | 1W |
| 120 | 5.60 | B | 1W |
| 124 | 5.62 | B | 1W |
| 128 | 5.64 | B | 1W |
| 132 | 5.66 | B | 1W |
| 136 | 5.68 | B | 1W |
| 140 | 5.70 | B | 1W |
| 149 | 5.745 | C | 4W |
| 153 | 5.765 | C | 4W |
| 157 | 5.785 | C | 4W |
| 161 | 5.805 | C | 4W |

.

Among these channels, there are eight widely used channels. These are divided into two called lower band UNII-1(36, 40, 44, and 48) and upper band UNII-3(149, 153, 157, and161). The remaining 15 channels are divided into lower UNII-2 having four channels, and upper UNII-2, which has 11 channels. The reason for the four-unit gaps is that the channels are 20MHz wide, while there is 5MHz spacing between one channel and the other. The upper bands uses 20 times as much power as the lower bands and the more power used to send signal, the likelihood of interference between networks [13].

### 3.2. IEEE808.11ac

In order to address the higher data rate throughput capacity require by the new technological growth in wireless applications, the IEE task group in 2008,developed an amendment to the IEEE802.11 PHY and MAC layer . IEEE802.11ac has an improving delivery capacity beyond its counterpart, IEEE802.11n.It is the next evolution of the Wi-Fi standard with the capacity to deliver multiple (High Definition) HD video streams simultaneously and provide improvements over 802.11n.It can cover 33m propagation distance by using 80MHz on the 5GHz band[28-30] This new technology proposed by  IEEE802.11- VHT ,aims at higher throughput of 1Gbps and with the frequency band of 6GHz or below excluding 2.4GHz frequency bandwidth of 80MHz and 160MHz option building on the 40MHz available in 802.11n.For spectrum efficiency, MU-MIMO technology is employed in this standard to achieve the aimed target throughput of 1Gbps.An obvious technical problem is envisaged in this technology. Due to its huge frequency bandwidth which  will only allow four frequency channels , will lead to severe inter cell interference due to frequency channel shortage among multiple basic service sets.TGac and TGad were established in Sep 2008 and Dec 2008 respectively, targeting completion of standardization late in 2012[31-33] The final approval of the IEEE802.11ac standard amendment is expected in December 2013, though as the moment, initial products with basic feature of 802.11ac has started to emerge on the market[30].

Table 2. Feature Enhancement Schemes Comparison: 802.11n/802.11ac [34]

|  | IEEE802.11n | IEEE802.11ac |
| --- | --- | --- |
| Frequency Band | 2.4 and 5 GHz | 5GHz only |
| Channel Widths | 20, 40 MHz | 20,40,80MHz, 106MHz optional |
| Spatial Streams | 1 to 4 | 1 to 8 total up to 4 per client |
| Multi- User MIMO | No | Yes |
| Single Stream (1x1) Maximum Client Data Rate | 150Mbps | 450Mbps |
| Three Stream (3x3) Maximum Client Data Rate | 450Mbps | 1.3Gbps |

## 3.3. Amendments to PHY and MAC layers for improved QoS

### 3.3.1. Higher Order Modulation Scheme

The maximum modulation in .11n is 64QAM constellation with six bits of coded information. This has been increased to 256QAM constellation with eight coded bits. 256QAM seems to be the current practical limit of digital modulation scheme, though efforts are on to increase to 512 and 1024 QAM. There are 256 possible states in this scheme, with each symbol representing eight bits. This makes this scheme very spectral efficient, the symbol rate is 1/8 of the bit rate. This is revealed in the fig.3.1, 33% increase is realized when 256QaM is used instead of the 64QAM. On the other hand, the constellation of this scheme are closely packed together, thus, more susceptible to noise and distortion, leading to higher bit error rate in higher modulation scheme [35, 36]

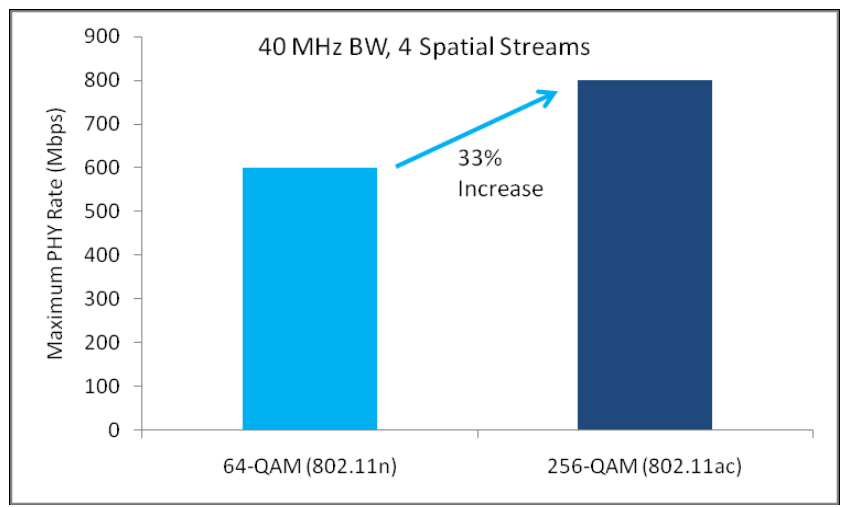

Figure 2. Performance enhancement in 802.11ac over 802.11n [37]

### 3.3.2. MIMO

MIMO (Multiple in Multiple out) techniques improves communication reliability as well as increase data throughput through spatial division multiplexing (SDM), without necessarily expanding the frequency band by using multiple antenna at the transmitter and receivers ends This results in signal interference. SVD-MIMO (singular value decomposition –multiple in multiple out) system can help in suppressing the signal interference as a result of lots of antenna by beamforming. This is effective in large scale MIMO systems [38, 39] while orthogonal division frequency multiplexing is a powerful tool to equalize received signals under multipath fading environments. The combination of these techniques is currently undergoing a great deal of attention in wireless communications[40, 41] Multiple-input multiple-output orthogonal frequency multiplexing (MIMO-OFDM) is a powerful technology that enhances communication capacity and reliance. This is being used in 802.11n WLAN system, defined as four spatial streams in spatial division multiplexing (SDM), it is expected that the upcoming standardization of 802.11ac will be defined in eight spatial streams in single-user MIMO (SU-MIMO[42]

### 3.3.3 Adaptive Beamforming Technology

This is one of the latest technologies in wireless communication system. It uses Smart Antenna techniques to produce multiple beams concentration to enhance signal of interest, and at the same time places nullity on the direction of interfering signal. The Beamformer varies a weight vector for this adaptive process with the ultimate goal to improve communication channel. The figure1b is the setup model for adaptive beamforming. The mathematical algorithms are shown in the following equations [25]

$$0 < \mu < 2 \frac{E[|u(n)|^2]}{E[|e(n)|^2]} D(n), \quad \text{Where}$$

$E[|e(n)|^2]$ is the power of the difference between the output and reference signals $\mu$ –step size
$E[|u(n)|^2]$ is power of incoming signal and $D(n)$ is the deviation
The error signal is given by;

$$e(n) = d(n) - \hat{w}^H(n)u(n) \tag{1}$$

$$\hat{w}(n+1) = \hat{w}(n) + \frac{\mu}{\|u(n)\|^2} u(n) e^*(n) \tag{2}$$

$$y(n) = w^H(n)x(n), \tag{3}$$

$$e(n) = d(n) - y(n) \tag{4}$$

Estimated of weight vector at time n+1,
$$w(n+1) = w(n) + \mu x(n) e^*(n)$$
$$= w(n) + \frac{\mu}{\|x(n)\|^2} x(n) e^*(n)$$

$$= w(n) + \frac{\mu}{\|x(n)\|^2} x(n) \|d(n) - {}^H_w x(n)\|^* \tag{5}$$

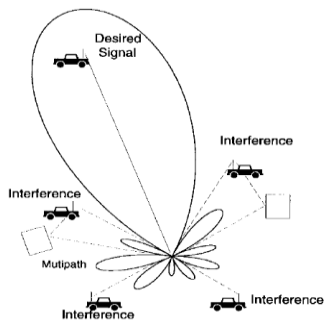
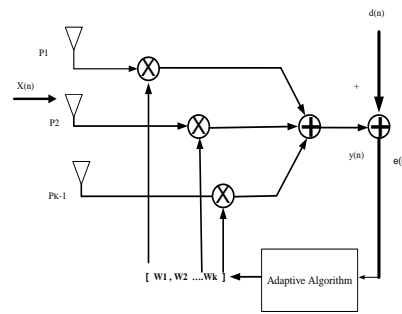

(6)

(a)             (b)

Figure 3. (a) Antenna Array Beamforming, (b) Block Diagram of Adaptive Beamforming [28]

# 4. METHODOLOGY

## 4.1. Empirical Measurements of wireless transmission on 2.4GHz channels

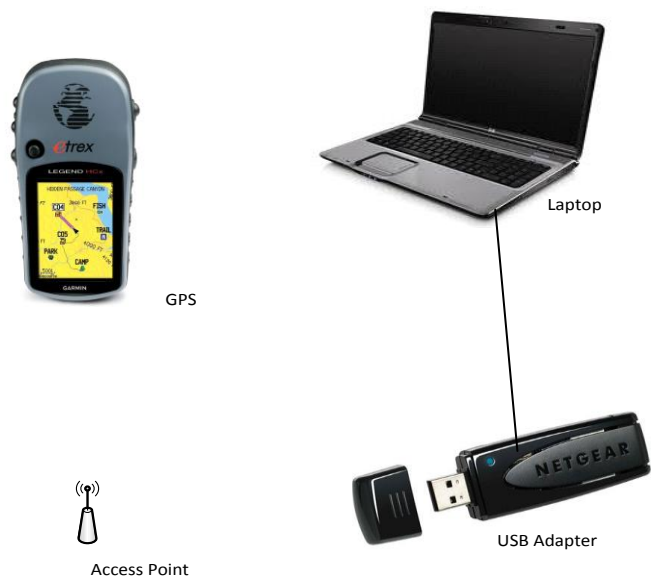

Figure 4. Microwave bands transmission experimental setup

The field measurements of radio signals at different locations from the various access points were taken with a net sniffer called Inssider, a 5GHz dual band dongle, GPS, and a laptop. Measurements were repeated about 25 times in each location for the different networks. While conducting the experiments, a huge amount of data was recorded in files and the data records include the time information (t), MAC address of AP, Signal Strength (SS) information, transmitter channel of AP, service set identifier (SSID).

The measured values were saved in kml file format from the personal computer and later extracted to Excel .Xml. Figs. 5-7 are the empirical models from the acquired data .Out of 23 non-overlapping channels in UNII band, only two were employed in the wireless network service of the University. They are channels 36 and 48, which also engaged in frequency reuse pattern. This band has higher data rate of 54Mbps hence higher throughput delivery compared to its unlicensed counterpart in the ISM band. The transmission distribution measurement of this band was also conducted and the simulations carried out. The following models are the outcomes.

## 4.2. Empirical Measurements of wireless transmission on IEEE802.11ac

To assess performance of 802.11ac, an ad-hoc system with the latest technology in wireless standards was used. The AirStation 1750 (Buffalo) is amongst the first routers to implement the newest WiFi standard: IEEE802.11ac. It promises throughput of 1300Mbps and also additional 450Mbps in the 802.11n radio.[30] The setup for the experiments are shown in figs 1&2 using WLI-4H-D1300 and WZR-D 1800H for the transmission using 5GHz channels, For comparison, in fig two, the client was changed to a WNDA 3100 N600 (Net gear) dual-band dongle using the 5GHz channels of 802.11n The measurement of throughputs over a range of distances up to 30m was observed on the PC connected to the wireless transmission link. The PCs were equipped with LANSpeedTest software for the monitoring of the throughputs.

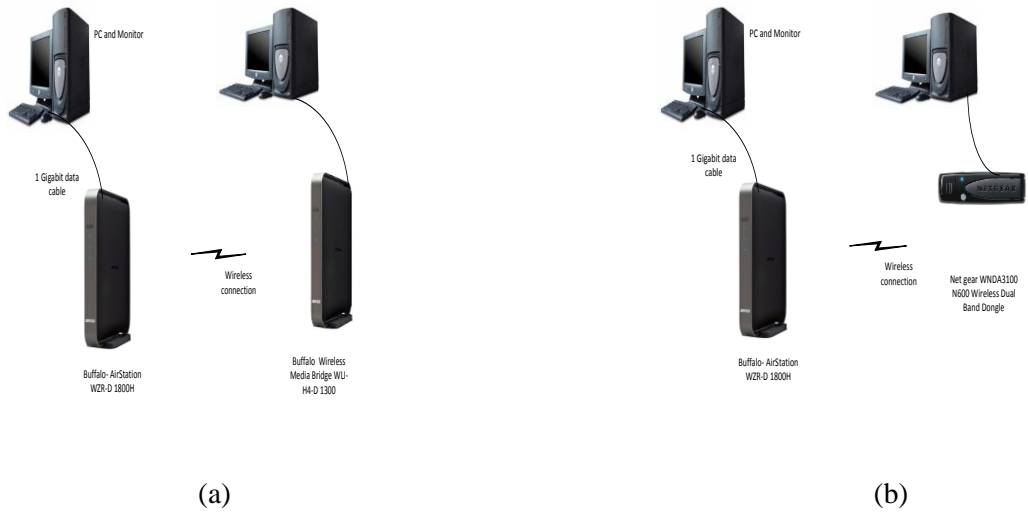

|        (a)        |        (b)        |

Figure 5. Buffalo AirStation AC1300/N900 Gigabit Dual Band Wireless Router and Client. (a) With IEEE802.11ac Standard, (b) with Net gear WNDA3100N600 Wireless Dongle Receiver on IEEE802.11n

### 4.3. Matlab Simulation-based evaluation of Beamforming in IEEE802.11ac

Table 4. Parameters for Simulation

| Carrier frequency | 5GHz |
|---|---|
| Sampling frequency | 2FC |
| Desires transmitted angle | Pi/4 |
| Interferer transmitted angle | Pi/4 |
| Step size | 0.32 |
| Number of element antenna array | 8 |

## 5. EMPIRICAL MODELS OF THE WIRELESS TRANSMISSION ON 2.4, 5GHz, IEEE802.11A, AND MATLAB SIMULATIONS

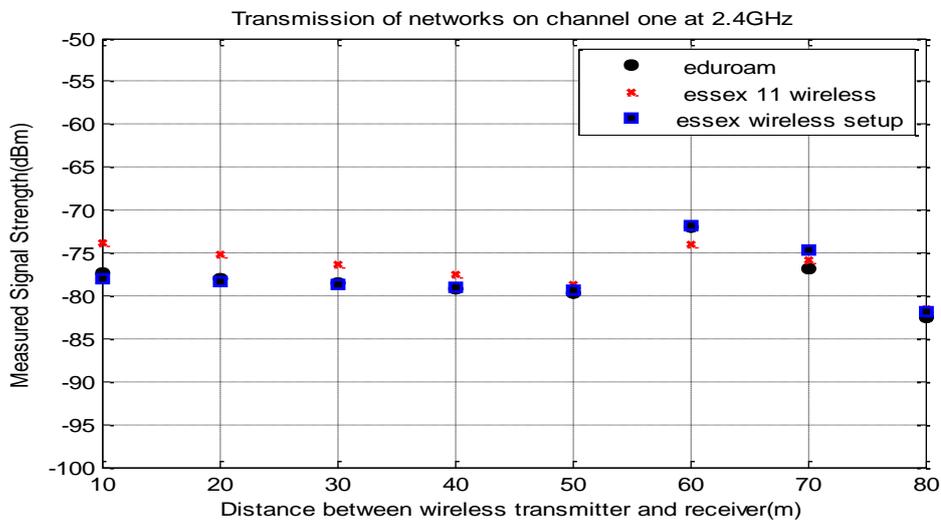

Figure 6. Channel One Transmission at 2.4GHz Coverage Distribution

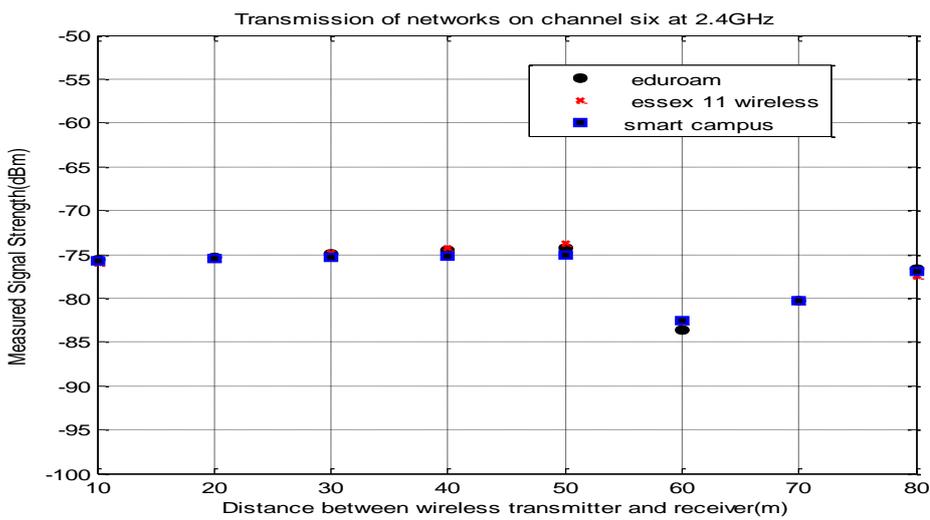

Figure 7. Channel Six Transmissions at 2.4GHz Coverage Distribution

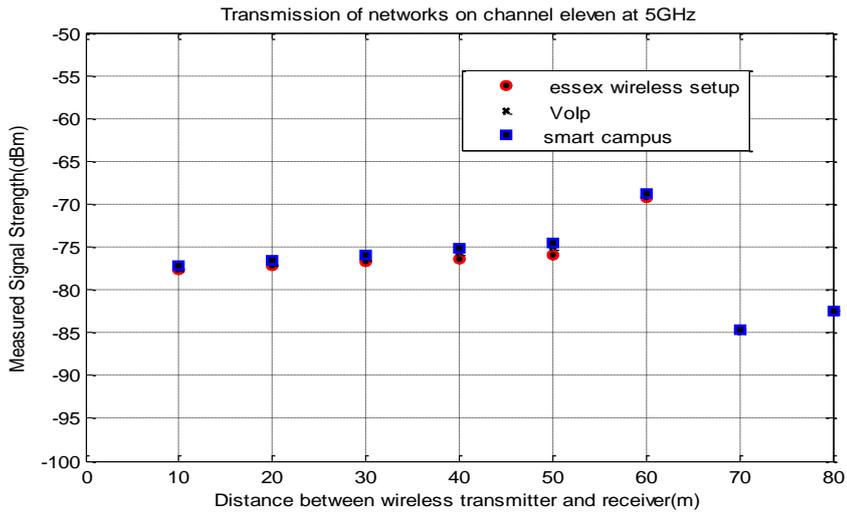

Figure 8. Channel Eleven Transmissions at 2.4GHz Coverage Distribution

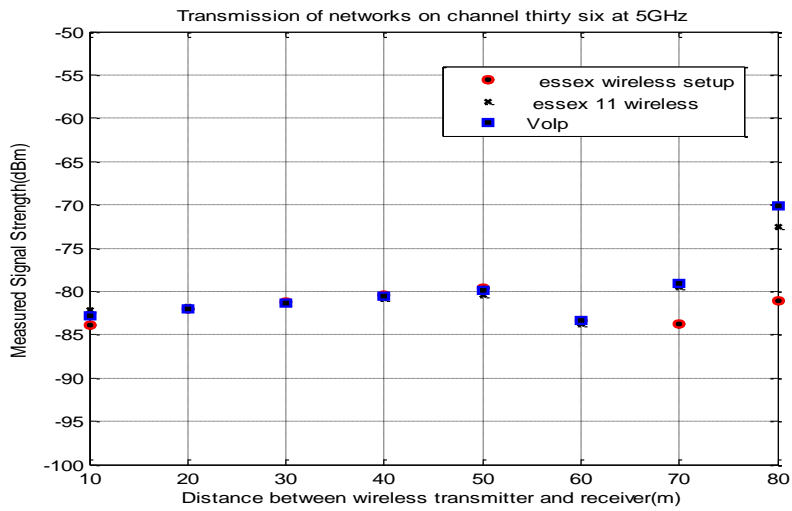

Figure 9. Channel 36 Transmissions at 5GHz Coverage Distribution

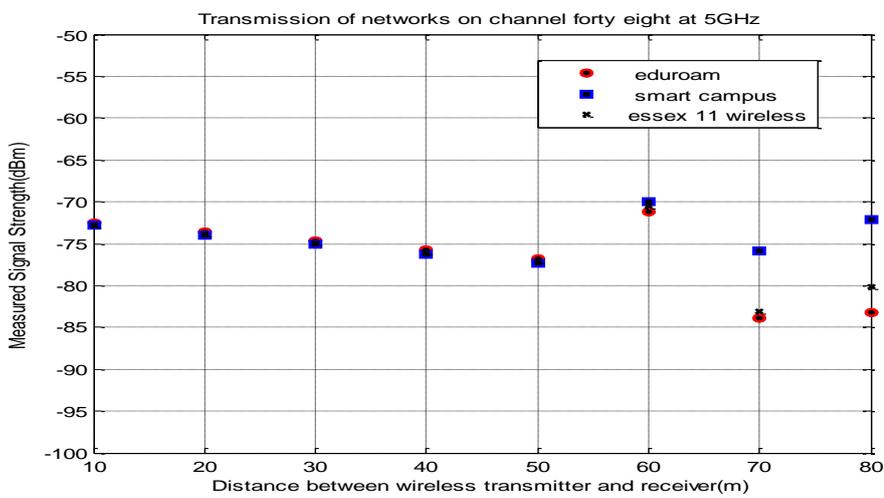

Figure 10. Channel 48 Transmissions at 5GHz Coverage Distribution

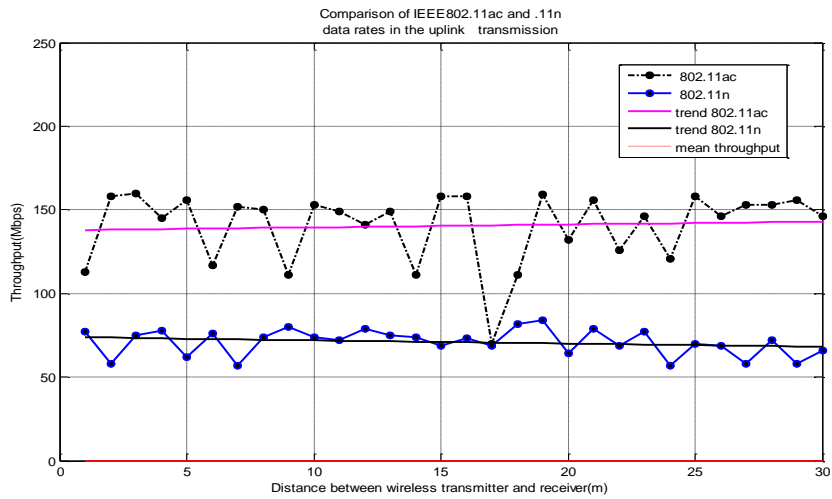

Figure 11.  Uplink Transmission Throughputs with AirStation Receiver and Dongle Receiver

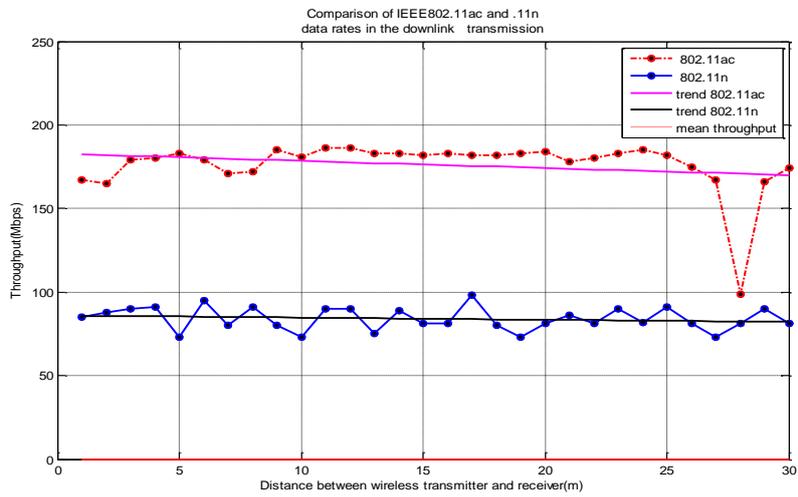

Figure 12.  Downlink Transmission Throughputs with AirStation Receiver and Dongle Receiver

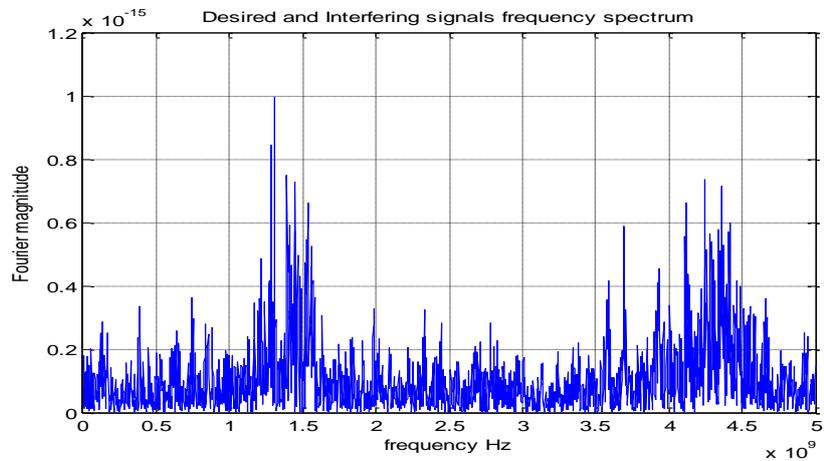

Figure 13.  Frequency Spectrum of Interfering and Desired Signals

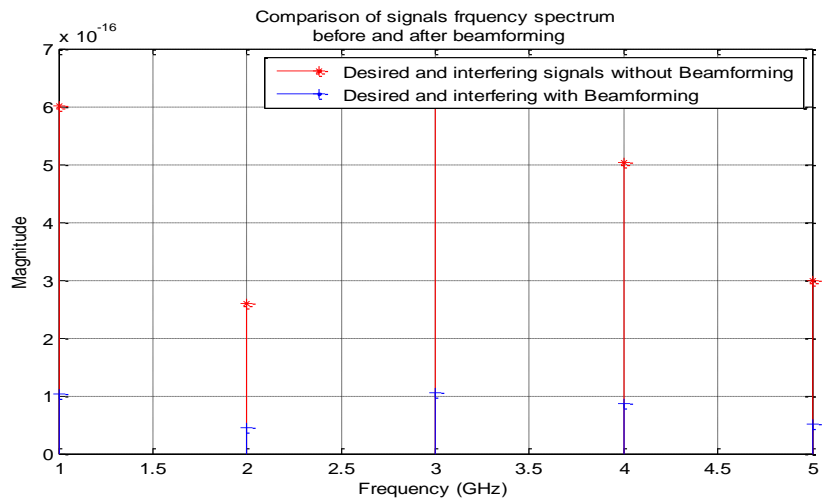

Figure 14. Frequency Spectrum before and after Beamforming Technique

## 5.2. Discussion

Measurements of transmission on microwaves bands were taken at twelve different locations in the University of Essex campus using the set up in figure 4.At each location, 25 signal strengths sampling from the access points (Aps) were recorded. Simulations of the measured values were carried out using the Excel.XML program, Haversine formula, and Frii's propagation law. The models are presented in figures 6-10. Some distance apart, show a measure of overlap. As shown in the figures, none of the access points was protected against interference. Each of the access points is degraded in throughput by certain percentage due to interference from other neighbouring access points. It is noted that eduroam and essex wireless setup access points were degraded by almost 100% in throughput.

The setup shown in figure 5 was used for wireless transmission on standard IEEE802.11ac deployed on frequency of 5GHz. The router was fixed in a position while the client/mobile terminal was moving at an equidistant points separated by 1m until a total coverage of 30m was reached in a laboratory hall at the University of Essex, where the measurement was conducted. The throughput levels received at each measure point is collected by a PC running with application software called LANSpeed Test software, as we moved along various points of measurement to monitor the throughput from the client. Figures 11-12 show the level of uplink and downlink throughputs of the Buffalo AirStation AC1300/N900 Gigabit Dual Band Wireless router and client when it is configured to work only in 802.11ac, and WNDA3100 N600 wireless dual-band dongle receiver when configured to work in 802.11n. As the figures show, the propagation square law was minimal and more revealed in IEEE802.11n configuration than IEEE802.11ac. There is appreciable measure of improved throughput in IEEE802.11ac over its counterpart IEEE802.11n. This established the benefits of beamforming technology in preserving the data to the targeted consumers. The matlab based simulation to investigate the effect of beamforming as a QoS enhancement scheme is revealed in figures13 and 14. The magnitude of the desired and interfering signals is shown in red before beamforming algorithms were implemented, while the blue colour signifies the magnitude of the desired signal after beamforming in figure 14. The drastic reduction in the magnitude is as a result of interference suppression through the shielding technique: beamforming technique, applied while the reception of desired signal is enhanced, hence improved quality of service.

## 6. CONCLUSION

The success in the growing wireless standards can be measured by the achievement of quality of service specifications by the designers. It becomes imperative to ascertain the performances of wireless devices. [31] Though. the theoretical peak data rates of 600Mbps and 1.3Gbps promised in IEEE802.11ac and IEEE802.11n respectively.[32] may not be realizable due to propagation impairment contending with wireless transmission in the real world scenario , nevertheless there is a considerable data throughput improvement on the counterpart 802.11n. The experiments have shown that IEEE802.11ac outperforms its lower counterpart IEEE802.11n in throughput delivery. This is as a result of the beamforming technology incorporated into this standard as a shielding technique for enhancement of system performance. The beamforming technology could be conceivably be modified to have better performance. The simulation based evaluation depicted in figures 13 and 14 revealed the effectiveness of beamforming technique.

## ACKNOWLEDGEMENTS

This work is sponsored by the Federal Government of Nigeria through the Tertiary Education Trust Fund (TETFUND) scheme.

## REFERENCES

[1] Q. Ni, L. Romdhani, and T. Turletti, "A survey of QoS enhancements for IEEE 802.11 wireless LAN," Wireless Communications and Mobile Computing, vol. 4, pp. 547-566, 2004.

[2] S. Chen, N. Ahmad, and L. Hanzo, "Smart beamforming for wireless communications: a novel minimum bit error rate approach," 2002.

[3] H. Liu, H. Darabi, P. Banerjee, and J. Liu, "Survey of wireless indoor positioning techniques and systems," Systems, Man, and Cybernetics, Part C: Applications and Reviews, IEEE Transactions on, vol. 37, pp. 1067-1080, 2007.

[4] K. K. Leung and B. J. Kim, "Frequency assignment for IEEE 802.11 wireless networks," in Vehicular Technology Conference, 2003. VTC 2003-Fall. 2003 IEEE 58th, 2003, pp. 1422-1426.

[5] D. C. K. Lee, M. J. Neve, and K. W. Sowerby, "The impact of structural shielding on the performance of wireless systems in a single-floor office building," Wireless Communications, IEEE Transactions on, vol. 6, pp. 1787-1695, 2007.

[6] A. Baid, "Multi-radio interference diagnosis in unlicensed bandsusing passive monitoring," Rutgers University-Graduate School-New Brunswick, 2011.

[7] K. Kim, "Interference mitigation in wireless communications," 2005.

[8] G. Ofori-Dwumfuo and S. Salakpi, "WiFi and WiMAX Deployment at the Ghana Ministry of Food and Agriculture," Research Journal of Applied Sciences, vol. 3, 2011.

[9] A. Sandeep, Y. Shreyas, S. Seth, R. Agarwal, and G. Sadashivappa, "Wireless Network Visualization and Indoor Empirical Propagation Model for a Campus WI-FI Network," World Academy of Science, Engineering and Technology, vol. 42, 2008.

[10] J. J. van Rensburg and B. Irwin, "Wireless Network Visualization Using Radio Propagation Modelling," 2005.

[11] Y. Lee, K. Kim, and Y. Choi, "Optimization of AP placement and channel assignment in wireless LANs," 2002, pp. 831-836.

[12]   A. Mishra, E. Rozner, S. Banerjee, and W. Arbaugh, "Exploiting partially overlapping channels in wireless networks: Turning a peril into an advantage," 2005, pp. 29-29.

[13]   S. Choi, K. Park, and C. Kim, "On the performance characteristics of WLANs: revisited," 2005, pp. 97-108.

[14]   D. Porcino and W. Hirt, "Ultra-wideband radio technology: potential and challenges ahead," Communications Magazine, IEEE, vol. 41, pp. 66-74, 2003.

[15]   X. Guo, S. Roy, and W. S. Conner, "Spatial reuse in wireless ad-hoc networks," 2003, pp. 1437-1442 Vol. 3.

[16]   S. Mare, D. Kotz, and A. Kumar, "Experimental validation of analytical performance models for IEEE 802.11 networks," in Communication Systems and Networks (COMSNETS), 2010 Second International Conference on, 2010, pp. 1-8.

[17]   A. Grilo and M. Nunes, "Performance evaluation of IEEE 802.11 e," in Personal, Indoor and Mobile Radio Communications, 2002. The 13th IEEE International Symposium on, 2002, pp. 511-517.

[18]   M. J. Ho, M. S. Rawles, M. Vrijkorte, and L. Fei, "RF challenges for 2.4 and 5 GHz WLAN deployment and design," 2002, pp. 783-788 vol. 2.

[19]   T. Murakami, R. Kudo, Y. Asai, T. Kumagai, and M. Mizoguchi, "Performance evaluation of distributed multi-cell beamforming for MU-MIMO systems," 2011, pp. 547-551.

[20]   C. Papathanasiou and L. Tassiulas, "Multicast Transmission over IEEE 802.11 n WLAN," in Communications, 2008. ICC'08. IEEE International Conference on, 2008, pp. 4943-4947.

[21]   Y. Lebrun, K. Zhao, S. Pollin, A. Bourdoux, F. Horlin, S. Du, and R. Lauwereins, "Beamforming techniques for enabling spatial-reuse in MCCA 802.11 s networks," EURASIP Journal on Wireless Communications and Networking, vol. 2011, pp. 1-13, 2011.

[22]   K. Fakih, J. F. Diouris, and G. Andrieux, "Analytical evaluation on the performance of ad hoc networks when using beamforming techniques," in Communications, 2008. ICC'08. IEEE International Conference on, 2008, pp. 2337-2342.

[23]   R. Vaze and R. W. Heath, "Transmission capacity of ad-hoc networks with multiple antennas using transmit stream adaptation and interference cancellation," Information Theory, IEEE Transactions on, vol. 58, pp. 780-792, 2012.

[24]   P. Xia, X. Qin, H. Niu, H. Singh, H. Shao, J. Oh, C. Y. Kweon, S. S. Kim, S. K. Yong, and C. Ngo, "Short range gigabit wireless communications systems: potentials, challenges and techniques," 2007, pp. 123-128.

[25]   K. M. alias Jeyanthi and A. Kabilan, "A Simple Adaptive Beamforming Algorithm with interference suppression," International Journal of Engineering and Technology, vol. 1, pp. 1793-8236, 2009.

[26]   S. M. Alamouti, "A simple transmit diversity technique for wireless communications," Selected Areas in Communications, IEEE Journal on, vol. 16, pp. 1451-1458, 1998.

[27]   N. V. Kajale, "Uwb and wlan coexistence: A comparison of interference reduction techniques," University of South Florida, 2005.

[28]   M. Matsuo, R. Ito, M. Kurosaki, B. Sai, Y. Kuroki, A. Miyazaki, and H. Ochi, "Wireless transmission of JPEG 2000 compressed video," 2011, pp. 1020-1024.

[29]   M. Matsuo, M. Kurosaki, Y. Nagao, B. Sai, Y. Kuroki, A. Miyazaki, and H. Ochi, "HDTV over MIMO wireless transmission system," 2011, pp. 701-702.

[30]   E. Perahia and M. X. Gong, "Gigabit wireless LANs: an overview of IEEE 802.11 ac and 802.11 ad," ACM SIGMOBILE Mobile Computing and Communications Review, vol. 15, pp. 23-33, 2011.

[31]   K. Nishikawa, "Ultra High-speed Radio Communication Systems and Their Applications-Current Status and Challenges."

[32] R. Liao, B. Bellalta, and M. Oliver, "DCF/USDMA: Enhanced DCF for Uplink SDMA Transmissions in WLANs."

[33] M. Park, "IEEE 802.11 ac: Dynamic Bandwidth Channel Access," 2011, pp. 1-5.

[34] R. Watson. (May 2012, Understanding the IEEE802.11ac Wi-Fi Standard. White Paper, 10.

[35] A. Technology. (2001, Digital Modulation in Communications systems An introduction, 48.

[36] Q. Incorporated. (2012, IEEE802.11ac: The Next Evolution of Wi-Fi Standards. 15.

[37] J. Montavont and T. Noel, "IEEE 802.11 handovers assisted by GPS information," 2006, pp. 166-172.

[38] K. Nishimori, N. Honma, T. Seki, and K. Hiraga, "On the Transmission Method for Short Range MIMO Communication," Vehicular Technology, IEEE Transactions on, pp. 1-1, 2011.

[39] T. Kaji, S. Yoshizawa, and Y. Miyanaga, "Development of an ASIP-based singular value decomposition processor in SVD-MIMO systems," 2011, pp. 1-5.

[40] S. Yoshizawa, S. Odagiri, Y. Asai, T. Gunji, T. Saito, and Y. Miyanaga, "Development and outdoor evaluation of an experimental platform in an 80-MHz bandwidth $2 \times 2$ MIMO-OFDM system at 5.2-GHz band," 2010, pp. 1049-1054.

[41] H. Kano, S. Yoshizawa, T. Gunji, T. Saito, and Y. Miyanaga, "Development of 600 Mbps $2 \times 2$ MIMO-OFDM baseband and RF transceiver at 5 GHz band," 2010, pp. 891-894.

[42] S. Yoshizawa, D. Nakagawa, N. Miyazaki, T. Kaji, and Y. Miyanaga, "LSI development of $8 \times 8$ single-user MIMO-OFDM for IEEE 802.11 ac WLANs," 2011, pp. 585-588.

**Femi-Jemilohun .O. Juliet** is a Ph.D. candidate at the School of Computer Science and Electronic Engineering the University of Essex, Colchester, United Kingdom .She became a student member of IEEE in 2013.She received her B.Eng. degree in Electrical and Electronics engineering, Ondo State University, Ado Ekiti, Nigeria in 1997 and her MEng degree from Federal University of Technology, Akure, Ondo State, Nigeria in 2010.

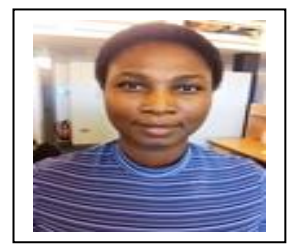

**Stuart D. Walker** received the B.Sc. (Hons.) degree in physics from Manchester University, Manchester, U.K., in 1973 and the M.Sc. and Ph.D. degrees in electronic systems engineering from the University of Essex, Essex, U.K., in 1975and 1981, respectively. After completing a period of contractual work for British Telecom Laboratories between 1979 and 1982, he joined the company as a Staff Member in 1982. He worked on various aspects of optical system design and wa spromoted to Head of the Regenerator Design Group in 1987. In 1988, he became Senior Lecturer and then Reader in Optical Communication at the University of Essex, U.K Currently, He is the Director of Postgraduate School, University of Essex.

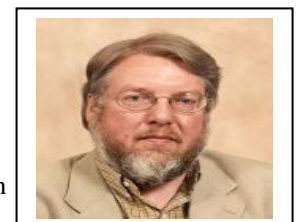